\documentclass[twocolumn,superscriptaddress,showpacs,preprintnumbers,amsmath,amssymb,prb]{revtex4}
\usepackage{amsmath}
\usepackage{subfig}
\usepackage{graphicx}
\begin{document}

\title{Thermodynamic stability of neutral Xe defects in diamond}

\author{D.~W.~Drumm}
\affiliation{School of Physics, University of Melbourne, Parkville 3010, Australia}

\author{M.~C.~Per}
\affiliation{Applied Physics, School of Applied Sciences, RMIT, Melbourne 3001, Australia}

\author{S.~P.~Russo}
\affiliation{Applied Physics, School of Applied Sciences, RMIT, Melbourne 3001, Australia}

\author{L.~C.~L.~Hollenberg}
\affiliation{School of Physics, University of Melbourne, Parkville 3010, Australia}

\date{\today}

\begin{abstract}
Optically active defect centers in diamond are of considerable interest, and \textit{ab initio} calculations have provided valuable insight into the physics of these systems.  Candidate structures for the Xe center in diamond, for which little structural information is known, are modeled using Density Functional Theory.  The relative thermodynamic stabilities were calculated for two likely structural arrangements.  The split-vacancy structure is found to be the most stable for all temperatures up to 1500K.  A vibrational analysis was also carried out, predicting Raman- and IR-active modes which may aid in distinguishing between center structures.
\end{abstract}


\maketitle

\section{Introduction}
There are over 500 optically active defects in diamond \cite{Davies94,Zaitsev00,Zaitsev01}, few of which have had their structures completely determined.  These defects' utilities range from potential bulk dopants to achieve superconductivity in diamond \cite{Ekimov04,Takano04} to bulk emissive properties in light-emitting diodes \cite{Zaitsev06}.  Other example defects include the neutral-vacancy GR1 \cite{Davies81}, nickel-based NIRIM-1 and -2 \cite{Isoya90}, boron acceptors \cite{Chrenko73}, as well as the only room-temperature stable single-photon source defects such as the well-studied nitrogen-vacancy center (NV) \cite{Kurtsiefer00}, the nickel-based NE8 center \cite{Gaebel04,Rabeau05}, silicon-vacancy (SiV) \cite{Wang06}, the unidentified 734 nm center \cite{Simpson09} and the possibly chromium-based MHz emission center \cite{Aharonovich09}.  In addition, the nitrogen-vacancy (NV) center has also been shown to act as a possible qubit for quantum computing \cite{Jelezko04}, as a magnetometer \cite{Chernobrod05,Balasubramanian08,Maze08,Balasubramanian09}, and to be useful in probing quantum decoherence \cite{Cole09,Hall09}.  It has also been used to demonstrate Quantum Key Distribution \cite{Beveratos02} using the BB84 scheme \cite{Bennett84}.  The optically-active Xe defect, the subject of this work, was only recently characterized \cite{Martinovich03,Martinovich04,Bergman07,Bergman09} and is of interest as a near-IR emitter with two zero phonon lines at 794 and 813 nm.
%

Optically active defects in diamond have high potential for technological applications, hence it is important to also consider their thermodynamic stability over the range of temperatures to which they are likely to be subjected, including during their creation.  Information on the relative stability of defects is also paramount in optimizing their manufacturing conditions.  Atomistic modeling of these types of defects offers several advantages: thermodynamic stabilities (both particular and relative) of defects can be evaluated, predictions of their vibrational spectrum made (which can aid in identifying the defect structure) and examination of potential electronic or optical properties may be undertaken.

The observation in Refs. \cite{Bergman07,Bergman09} of a Xe-based defect in diamond emitting at 794 and 813 nm makes the case for detailed \textit{ab initio} studies of such centers even stronger.
The site symmetry of that defect was established as trigonal by their polarization study.  This defect is likely to be thermodynamically driven due to the experimental conditions under which it is formed (180-500 keV implantation with 800-1400$^{\circ}$C annealing), but little else has been published upon Xe-related defects in diamond other than recent studies indicating that Xe chemically reacts with the diamond lattice \cite{Goss09}, and that the defect is comprised of one Xe atom (with assorted vacancies) \cite{Bergman09}.

Density Functional Theory (DFT) \cite{Hohenberg64,Kohn65a} has provided important insight into the nature of such defects \cite{Goss96,Schultz06,Hossain08,Goss09} and is excellent at predicting ground-state properties.  The stability of various morphologies \cite{Barnard03a},  modeling of dopants \cite{Barnard03c,Barnard03d,Hossain08}, and semiconductor band-gap modification \cite{Barnard03g} have been characterised.

As a material, diamond offers several other attractive properties, including the largest optical transparency window, Young's modulus, and the advantage of a chemically inert host lattice \cite{Jelezko06}.  The combination of this exceptional host material with its plethora of dopants offers much scope for investigation and development.  The lack of definitive structure models for many of these defect centers requires attention.

In this paper, as the true geometric structure of the defect has not been identified by experiment, we use DFT to study the thermodynamic stability of neutral Xe-vacancy defects in diamond which both have the same symmetry as those observed by Bergman \textit{et al.} \cite{Bergman07,Bergman09}, and contain one Xe atom \cite{Bergman09}.  We specifically consider a substitutional Xe atom with either one (Xe$_{\text{s}}$--V) or three (Xe$_{\text{s}}$--3V) vacant lattice sites directly adjacent to the Xe atom.
%
%
The paper is organised as follows: in Section \ref{sec:theory} we discuss defect formation free energies and how they may be compared; Sec. \ref{sec:method} details the DFT parameters used in calculations; Sec. \ref{sec:structure} presents relaxed defect structures; in Sec. \ref{sec:vDOS} vibrational results are discussed; the relative defect formation free energy is shown in Sec. \ref{sec:energies}, and conclusions are drawn in Sec. \ref{sec:conclusions}.

\section{Defect Formation Energy \& Thermodynamic Stability}
\label{sec:theory}
To compare formation free energies between chemical structures requires accounting for the changes one must make to transform from one structure to the other.  The defect formation free energy ($\Delta F_{f}$) is the free energy required to form the defect system from a host (defect-free) lattice $H$.  For a defect system $X$ consisting of dopant atoms and associated lattice vacancies, which are formed by the removal of $n_{i}$ atoms of chemical species $i$, $\Delta F_{f}$ becomes (for any temperature $T$ within the harmonic approximation):
\begin{equation}
\label{eq:deltaff}
\Delta F_{f}=F\left(X\right)-\left[F\left(H\right)+\sum_{D}n_{D}\mu_{D}-\sum_{i}n_{i}\mu_{i}\right]+cct.
\end{equation}
Here $F\left(X\right)$ is the total free energy for the defected system, and $F\left(H\right)$ the total free energy of the defect-free host lattice.  The chemical potentials of the dopant atom ($D$) and the atom(s) of chemical species $i$ removed to create a vacancy in the defect lattice are $\mu_{D}$ and $\mu_{i}$.  The number of the dopant atoms inserted is $n_{D}$; $n_{i}$ is the number of lattice atoms removed to create vacancies.  Finally, $cct$ corresponds to charge correction terms which are added to equation \ref{eq:deltaff} if the atoms removed or added have a charge \cite{Goss09,Goss07}.

The free-energies are defined as
\begin{equation}
\label{eq:Fx}
F\left(X\right)=E\left(X\right)+F_{\text{vib}}\left(X\right)-TS_{\text{config}},
\end{equation}
\begin{equation}
\label{eq:Fh}
F\left(H\right)=E\left(H\right)+F_{\text{vib}}\left(H\right),
\end{equation}
where $E\left(X\right)$ is the total (DFT) energy for the relaxed defect system, and $E\left(H\right)$ is the total (DFT) energy of the relaxed defect-free host lattice.  $F_{\text{vib}}$ is the vibrational free energy contribution from the system $X$ or host $H$ as given by
\begin{multline}
\label{eq:Fvib}
F_{\text{vib}}=k_{B}T\int_{0}^{\infty}g\left(\omega\right)\text{ln}\left[2~\text{sinh}\left(\frac{\hbar\omega}{2k_{B}T}\right)\right]d\omega,
\end{multline}
where $g(\omega)$ is the (harmonic approximation) phonon density of states for system $X$ or $H$. It is important to note that here $g(\omega)$ has been normalized such that
\begin{equation}
\label{eq:Np}
\int_{0}^{\infty}g\left(\omega\right)d\omega=N_{P}.
\end{equation}
Here $N_{P}$ is the number of phonon branches in the supercell.  As written, $F_{\text{vib}}$ also contains the zero-point energy $E_{\text{ZP}}=\int_{0}^{\infty}g\left(\omega\right)\frac{1}{2}\hbar\omega d\omega$ contribution of the $X$ or $H$ system, viz.,
\begin{multline}
F_{\text{vib}}=\int_{0}^{\infty}g\left(\omega\right)k_{B}T~\text{ln}\left[2~\text{sinh}\left(\frac{\hbar\omega}{2k_{B}T}\right)\right]d\omega \\
=E_{\text{ZP}}-k_{B}T\int_{0}^{\infty}g\left(\omega\right)\text{ln}\left[1-e^{-\frac{\hbar\omega}{k_{B}T}}\right]d\omega.
\end{multline}
The configurational entropy contribution $S_{\text{config}}$ is given by $S_{\text{config}} = k_{B}~$ln$\left(\Omega_{X}\right)$, where $\Omega_{X}$ enumerates the possible orientations of the defect ($X$) within the supercell.  In both cases, our defects require the Xe atom to be on a lattice site (initially), and have a distinct orientation towards one of the nearest-neighbor lattice sites.  This evaluates as:
\begin{equation}
\Omega_{X}=n_{l}n_{o}=864,
\end{equation}
where $n_{l}$ is the number of lattice sites in the supercell (216) and $n_{o}$ is the number of distinct orientations the Xe defect may have per lattice site (or number of nearest-neighbor C atoms, 4).  While this works for Xe--3V, for Xe--V we must, as the final structure is significantly different from the initial (see Sec. \ref{sec:structure}), include a term to avoid double-counting:
\begin{equation}
\Omega_{\text{Xe--V}}=\frac{n_{l}n_{o}}{2}=432.
\end{equation}

The chemical potential of the carbon atoms removed is
\begin{equation}
\label{eq:chempotC}
\mu_{\text{C}}=\frac{E_{DFT}\left(H\right)+F_{\text{vib}}\left(H\right)}{n},
\end{equation}
where $n$ is the number of C atoms in the defect-free supercell of diamond.  For this work we consider the formation of a neutral dopant (Xe atom) in a covalent system (diamond), thus charge correction terms may be ignored.

As we are considering two defected systems (Xe--V and Xe--3V), in determining the relative thermodynamic stability of the two defect configurations we need only consider the relative defect formation free energy
\begin{equation}
\Delta F_{f}^{\text{REL}}=\Delta F_{f}\left(\text{Xe--V}\right)-\Delta F_{f}\left(\text{Xe--3V}\right)
\end{equation}
allowing us to divorce common bulk properties from relative stability calculations:
\begin{equation}
\label{eq:deltaffrel}
\Delta F_{f}^{\text{REL}}=F\left(\text{Xe--V}\right)-\left[F\left(\text{Xe--3V}\right)+2\mu_{\text{C}}\right].
\end{equation}
As defined, $\Delta F_{f}^{\text{REL}}>0$ means that the Xe--3V case is more stable than the Xe--V case.

Note that the $\mu_{\text{Xe}}$ term does not appear in the above expression, due to both our defects being comprised of one Xe dopant atom, and the $\mu_{\text{Xe}}$ terms cancelling.

\section{Methodology}
\label{sec:method}
All DFT calculations were performed using the \textit{ab initio} total-energy and molecular dynamics package \textsc{vasp} (Vienna \textit{ab initio} simulation package) developed at the Universit\"at Wien \cite{Kresse96}.  We used the generalized-gradient approximation (GGA), with the PBE exchange-correlation functional \cite{Perdew96}, and Plane Augmented Wave (PAW) \cite{Kresse99} potentials for C and Xe. The plane-wave cut-off was set at 400 eV for all calculations. A 4$\times$4$\times$4 Monkhorst-Pack \textit{k}-point mesh was used as this has been found to give accurate representation of the geometry and electronic structure \cite{Barnard02, Barnard03h, Hossain08, Goss09}.  The bulk diamond lattice constant was checked using these parameters, found to be 3.562 \r{A}, and this value was used for all subsequent calculations.

All calculations were performed by constructing a 3$\times$3$\times$3 simple cubic supercell of diamond (216 atoms), removing the appropriate number of carbon atoms to create vacancies and inserting Xe into a substitutional lattice site.  The vacancy sites were always in first nearest neighbor positions to the Xe atom. For the defect diamond lattices, the Xe-vacancy defects were constructed such that they had trigonal crystallographic point group symmetry in accordance with experimental data \cite{Bergman07}.

In performing the geometry optimization the relaxation was constrained so that the supercell remained cubic in shape. After relaxation the force on a C atom was typically less than 0.001 eV/\r{A} and in all cases the trigonal symmetry of the defect remained.

%

\section{Defect Geometric Structure}
\label{sec:structure}
\subsection{Split-vacancy interstitial Xe}
\label{sec:1V}
\begin{figure}[ht]
\mbox{\subfloat{\label{fig:XeVstart}\includegraphics[width=0.8\linewidth]{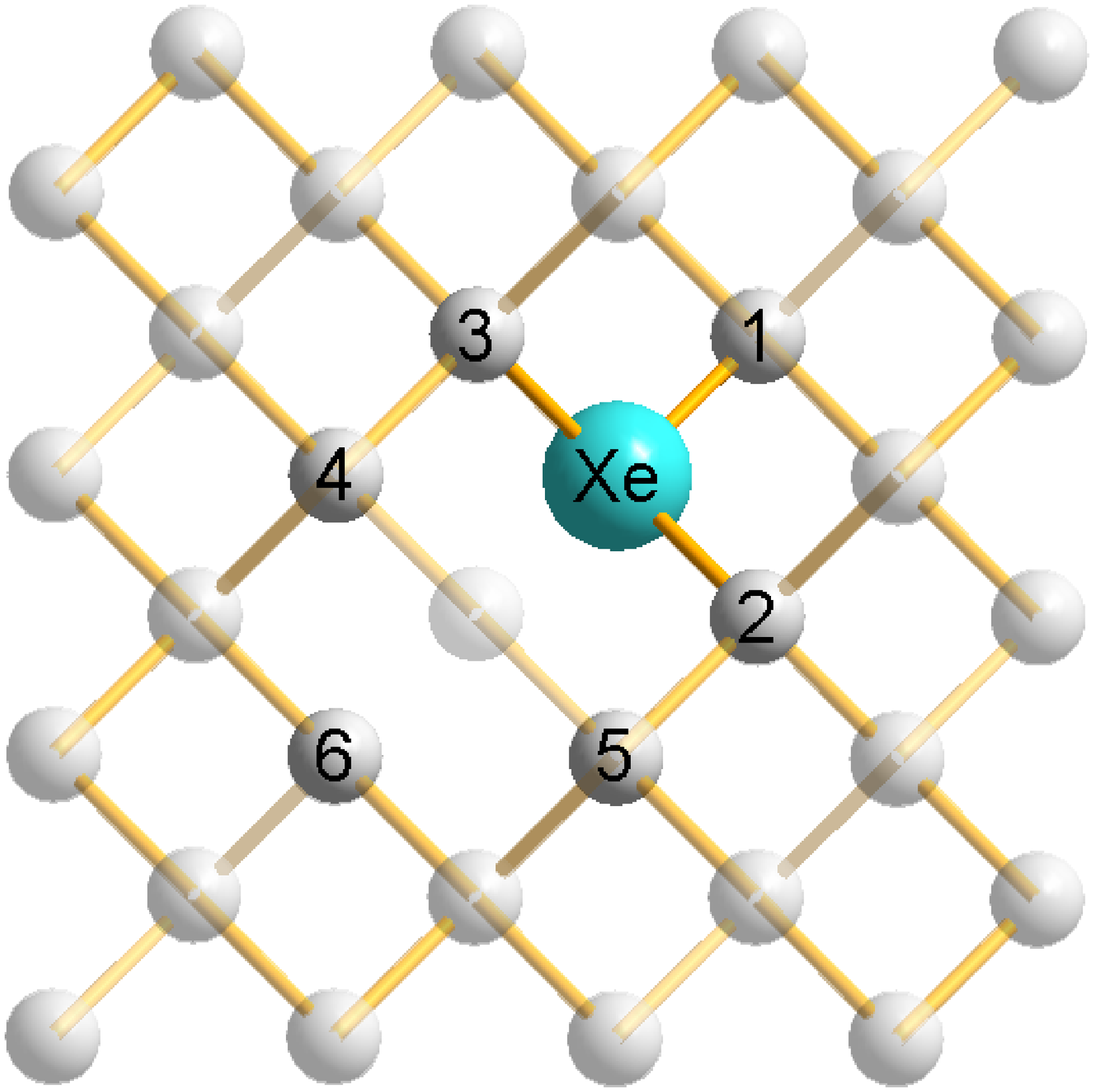}}}
\mbox{\subfloat{\label{fig:XeVend}\includegraphics[width=0.8\linewidth]{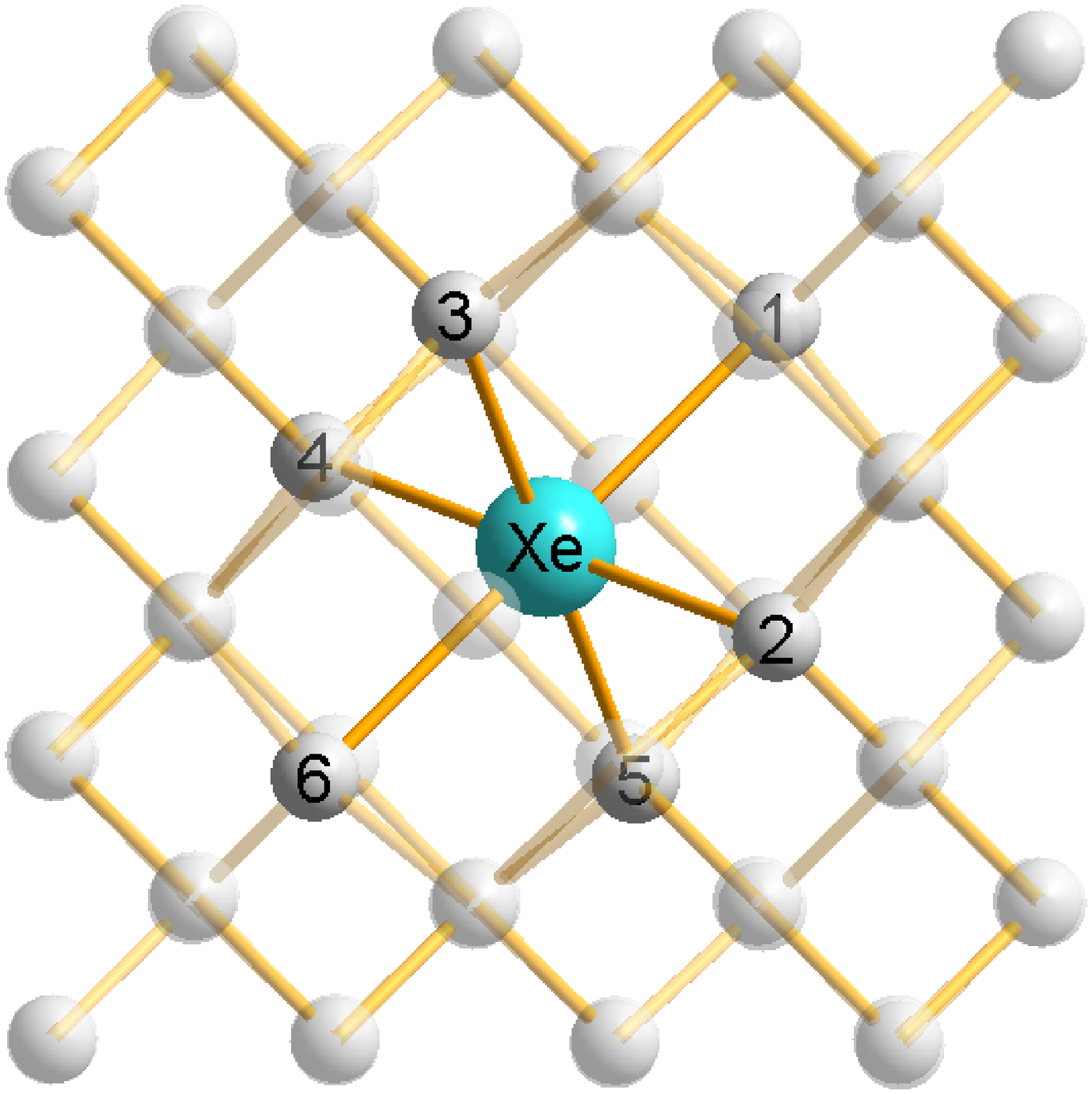}}}
\caption{(Color online)  Split-vacancy Xe structure, viewed from $\left[100\right]$ direction.  C atoms shown in grey, nearest neighbors solid grey and numbered, Xe atom shown larger in blue.  (a) Initial structure.  (b) Relaxed structure.}
\end{figure}

Upon relaxation the initial Xe-vacancy geometry (Xe$_{\text{s}}$--$V$ in Fig. \ref{fig:XeVstart}), with all atoms occupying C lattice sites, formed a split-vacancy interstitial structure with the site symmetry of the Xe atom being $C_{3v}$ (Fig. \ref{fig:XeVend}).  The Xe atom was then found to be at the centre of the divacancy, where it has $D_{3d}$ symmetry (still within the constraint of trigonal symmetry).  In the relaxed structure, the Xe atom is six-fold coordinated with the nearest-neighbour C atoms, with bond lengths of 2.15 \r{A}.  The bonds form two distinct sets of three (C$_{1-3}$ and C$_{4-6}$); bond angles within sets are different from those between sets.  More details may be seen in Table \ref{tab:1V}.  This structure is commensurate with the Si--$V$ defect in diamond studied by Goss \textit{et al.} \cite{Goss96}

\begin{table}
  \begin{center}
    \begin{tabular}{c@{\extracolsep{0.1\linewidth}}c@{\extracolsep{0.1\linewidth}}c}
      \hline
      \hline
      Bond Length																												& Relaxed  								& Change from\\
      / Angle																														& amount									& initial\\
      \hline
			$\text{C}_{i}\text{--Xe}$ 																				& 2.15$ \text{\r{A}}$  	  & +39.6\%\\
			\hline
			$\angle \text{C}_{1\text{--}3}\text{--Xe--C}_{1\text{--}3}$	      & 84.436$^{\circ}$				& -22.9\%\\
			$\angle \text{C}_{4\text{--}6}\text{--Xe--C}_{4\text{--}6}$	      & 84.316$^{\circ}$				& +40.5\%\\
			$\angle \text{C}_{1\text{--}3}\text{--Xe--C}_{4\text{--}6}$	      & 95.640$^{\circ}$				& + 6.27\%\\
      \hline
      \hline
    \end{tabular}
  \end{center}
  \caption{Bond lengths and angles for relaxed split-vacancy Xe}
  \label{tab:1V}
\end{table}

\subsection{Three-vacancy defect (Xe$_{\text{s}}$--3V)}
\label{sec:3V}
\begin{figure}[ht]
\mbox{\subfloat{\label{fig:Xe3Vstart}\includegraphics[width=0.8\linewidth]{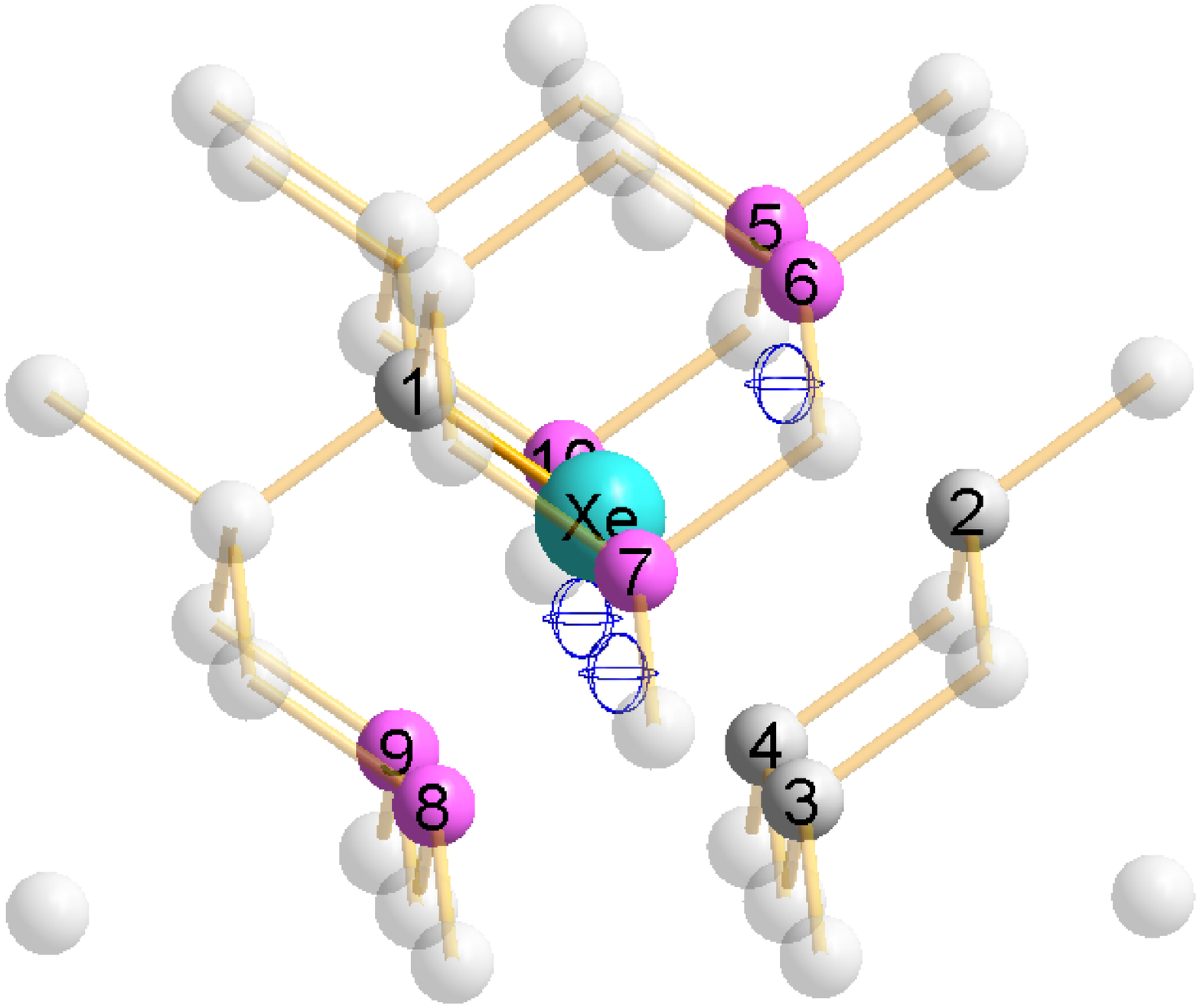}}}
\mbox{\subfloat{\label{fig:Xe3Vend}\includegraphics[width=0.8\linewidth]{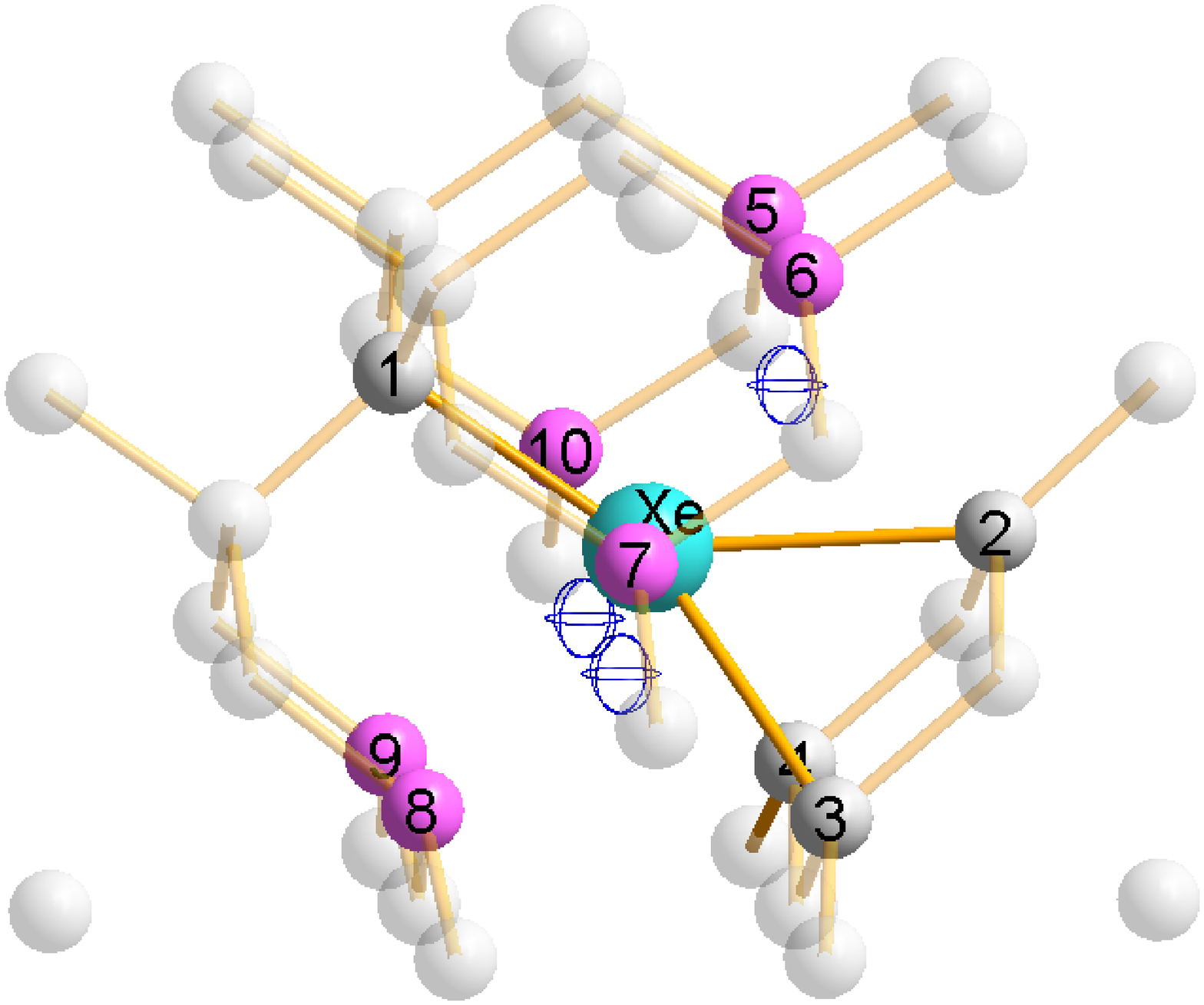}}}
\caption{(Color online)  Xe$_{\text{s}}$--3V, viewed from approximately the $\left[ 1 1 0 \right]$ direction.  C atoms shown in grey, nearest neighbors solid grey and numbered, 6-member C ring shown in pink and numbered, Xe atom shown larger in blue, vacant sites shown as blue wire ellipse frames.  (a) Initial structure.  (b) Relaxed structure.}
\end{figure}

The three-vacancy case (initial geometry shown in Fig. \ref{fig:Xe3Vstart}) was also considered.  Initially, the Xe is left with one nearest-neighbor C atom (C$_{1}$), and three directly adjacent vacant sites.  Of the surrounding atoms, nine (C$_{2\text{--}10}$) are also adjacent to the vacancies.  Six of these (C$_{5\text{--}10}$) form a coplanar hexagon about the Xe atom.

The relaxed geometry (Fig. \ref{fig:Xe3Vend}) shows that the Xe atom migrates 0.38 \r{A} (out of the hexagon plane) to an interstitial position between the four vacant lattice sites (one being the initial position of the Xe atom) with the site symmetry of the Xe atom being $C_{3v}$.  C$_{1-4}$ move outward by 0.22 \r{A}.  The hexagon distorts slightly, with pairs of atoms expanding outwards (along their perpendicular bisector) by 8.66 \r{A}, maintaining $C_{3v}$ symmetry.  More details are located in Table \ref{tab:3V}.
\begin{table}
  \begin{center}
    \begin{tabular}{c@{\extracolsep{0.05\linewidth}}c@{\extracolsep{0.05\linewidth}}c}
      \hline
      \hline
      Bond Length /  									     																																				& Relaxed				   				& Change from \\
      Distance / Angle																																														& amount									& initial \\
      \hline
			$\text{C}_{1}\text{--Xe}$       																																						& 2.13$ \text{\r{A}}$  	  & +38.3\%\\
			$\text{C}_{2\text{--}4}\text{--Xe}$ 																																				& 2.39$ \text{\r{A}}$  	  & - 5.28\%\\
			$\text{C}_{5\text{--}10}\text{--Xe}$																																				& 2.63$ \text{\r{A}}$  	  & + 4.47\%\\
			\hline
			$\text{C}_{5}\text{--C}_{6}$ $\left(7\text{--}8,9\text{--}10\right)$ 																				& 2.67$ \text{\r{A}}$  	  & + 5.86\%\\
			$\text{C}_{6}\text{--C}_{7}$ $\left(8\text{--}9,10\text{--}5\right)$																				& 2.52$ \text{\r{A}}$  	  & + 0.11\%\\
			\hline
			$\angle \text{C}_{1}\text{--Xe--C}_{5\text{--}10}$					      																					& 80.334$^{\circ}$				& -10.7\%\\
			$\angle \text{C}_{5}\text{--Xe--C}_{6}$ $\left(7\text{--Xe--}8,9\text{--Xe--}10\right)$	      							& 60.878$^{\circ}$				& + 1.46\%\\
			$\angle \text{C}_{6}\text{--Xe--C}_{7}$ $\left(8\text{--Xe--}9,10\text{--Xe--}5\right)$	      							& 57.247$^{\circ}$				& - 4.59\%\\
			$\angle \text{C}_{2\text{--}4}\text{--Xe--C}_{5\text{--}10}$	      																				& 67.828$^{\circ}$				& +13.0\%\\
			$\angle \text{C}_{2\text{--}4}\text{--Xe--C}_{2\text{--}4}$	      																					& 63.129$^{\circ}$				& + 5.23\%\\
      \hline
      \hline
    \end{tabular}
  \end{center}
  \caption{Bond lengths and angles for relaxed Xe--3V}
  \label{tab:3V}
\end{table}
\section{Defect formation energies}
\label{sec:energies}
\begin{figure}[ht]
\includegraphics[width=\linewidth]{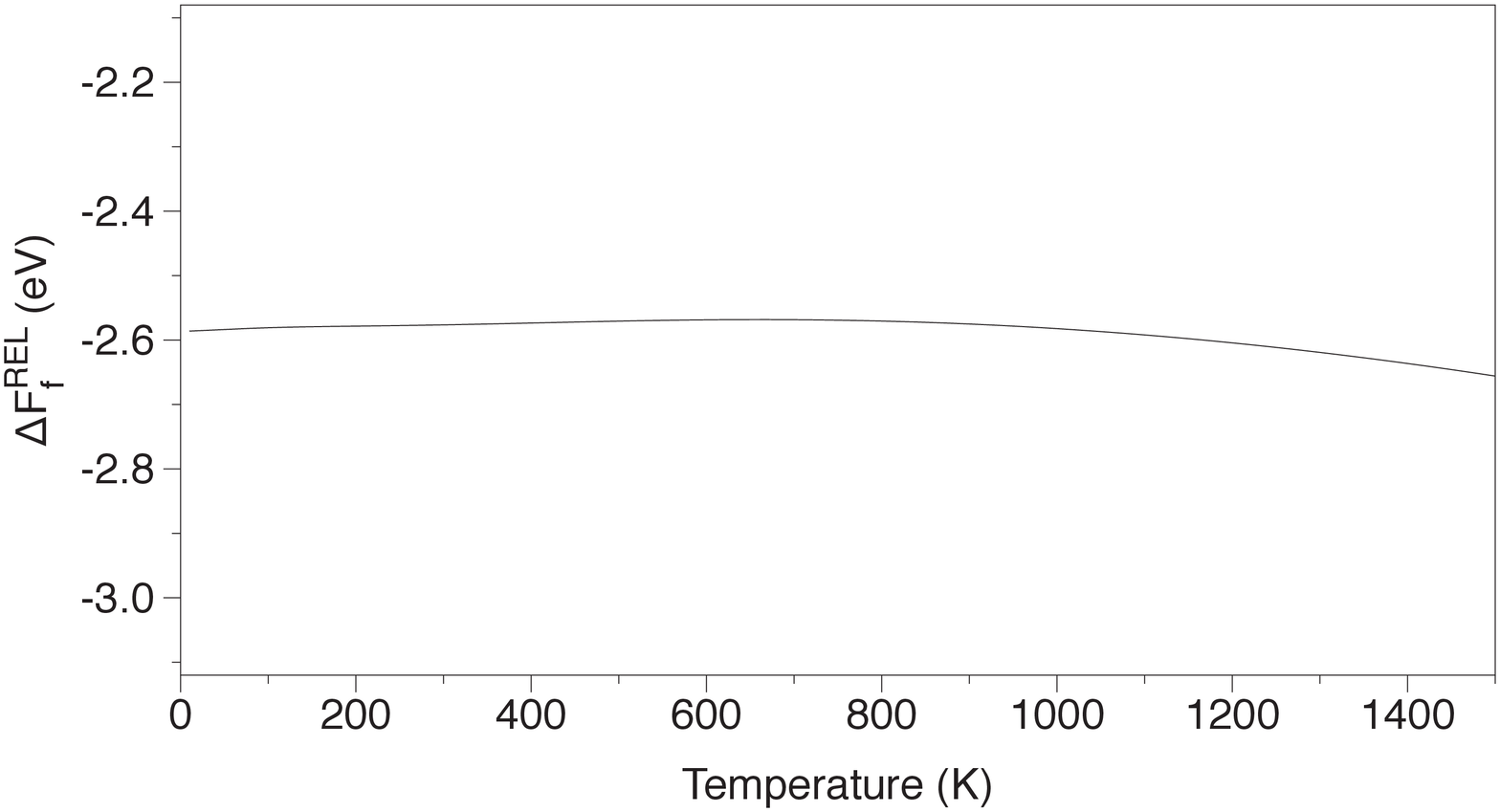}
\caption{Relative defect free energy of formation; $F\left(\right.$Xe--V$\left.\right)-\left[F\left(\right.\right.$Xe--3V$\left.\left.\right)+2\mu_{\text{C}}\right]$.  (Color online)}
\label{fig:deltaffrel2}
\end{figure}

\begin{figure}[ht]
\includegraphics[width=\linewidth]{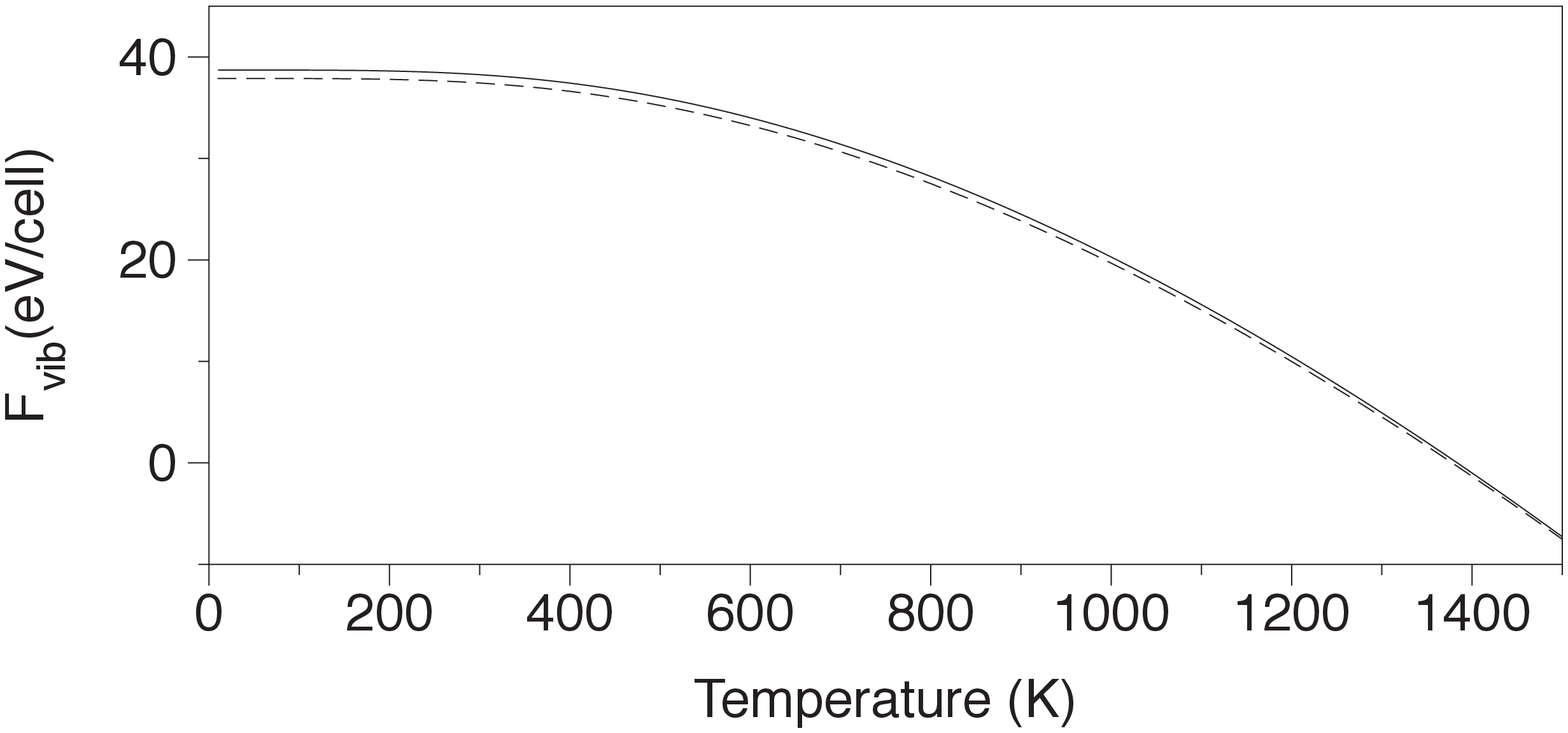}
\caption{Vibrational contribution to the free energy.  Split-vacancy Xe defect (solid line), Xe--3V defect (dashed line).  (Color online)}
\label{fig:FvibvsT}
\end{figure}

Using the equations detailed in Section \ref{sec:theory}, the relative defect formation free energy ($\Delta F_{f}^{\text{REL}}$ from Eq. \ref{eq:deltaffrel}) was calculated over temperatures ranging from 0-1500K and can be seen in Fig. \ref{fig:deltaffrel2}.  The split-vacancy defect is 2.59 eV more stable than the Xe--3V defect at all temperatures considered.  The variation with temperature is small, about 3\% of the total value, indicating that the difference in their behaviour with respect to temperature is small compared to the other terms in $\Delta F_{f}^{\text{REL}}$ and that for the purpose of determining the qualitative stability ordering temperature effects are irrelevant.  The relative stability of the split-vacancy defect is slightly enhanced at high temperatures.  Neglecting $E_{\text{ZP}}$ (which comes at significant computational expense compared to a simple DFT calculation), leads to a relative zero-temperature defect formation energy of -3.06 eV, overestimating the stability by $\approx$18\%.
Finally, in Fig. \ref{fig:FvibvsT} we show the vibrational free energy contribution to the total energy as a function of temperature for each defect.  The Xe--3V case (dashed line) is vibrationally more stable at every temperature studied.  At 0K, the difference is 0.83 eV, and at 1500K, 0.25 eV.  We may therefore infer that the energetic expense of removing two C atoms from the diamond lattice outweighs the steric benefit of having the extra space for the Xe atom and unpaired C electrons to inhabit.
\section{Vibrational DOS}
\label{sec:vDOS}
The vibrational properties of the systems were then calculated using lattice dynamics under the harmonic approximation.  The derivatives of the forces required to build the dynamical matrix were obtained using a finite-difference scheme in which individual nuclei were displaced by 0.04 \r{A} from their equilibrium positions and the resulting forces on all the nuclei calculated.  Symmetry was used to reduce the number of displacements required to build the full dynamical matrix.  The calculated forces were used in the program \textsc{phon} \cite{Alfe09} to construct and diagonalise the dynamical matrix to obtain the vibrational Density of States (vDOS) and the thermodynamic properties of the systems.

\begin{figure}[ht]
\mbox{\subfloat{\label{fig:XeVvdosBZ}\includegraphics[width=0.9\linewidth]{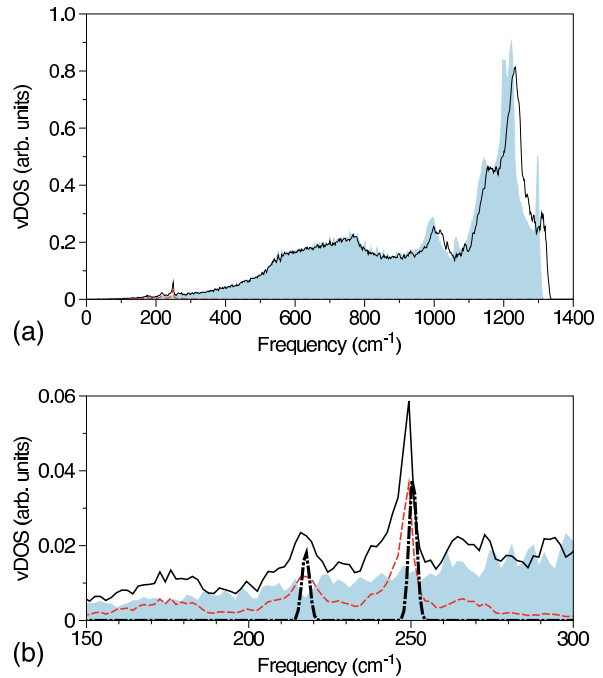}}}
\caption{(Color online)  Vibrational density of states for the relaxed split-vacancy Xe defect.  Total vDOS (solid black line), Xe-projected vDOS (dashed red line), Gaussian-smeared Xe-projected vDOS at $\Gamma$ (dot-dashed green line), bulk diamond vDOS (shaded background).  (a) Full calculated vDOS.  (b) Region where Xe-projected vDOS is significant.}
\label{fig:XeVvDOS}
\end{figure}

\begin{figure}[ht]
\includegraphics[width=0.67\linewidth]{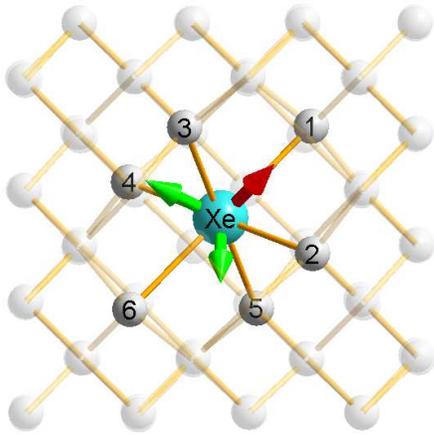}
\caption{Displacements of the Xe atom under the two strongly Xe-localized vibrational modes in the split-vacancy Xe defect.  A$_{1}$ mode (red arrow), doubly-degenerate E mode (green arrow).  (Color online)}
\label{fig:vec}
\end{figure}

Figure \ref{fig:XeVvdosBZ} shows the total vDOS of bulk diamond (shaded) and for the Xe split-vacancy (black), averaged over the entire Brillouin zone.  The region where the contribution from the Xe atom, the Xe-projected vDOS (dashed red) is significant is shown in Fig. 5b.  Three strong Xe-localized modes are extant at the $\Gamma$ point (dot-dashed green).  As the site symmetry of the Xe atom and the crystallographic point group symmetry of the relaxed lattice are identical ($C_{3v}$), the vibrational modes involving the Xe atom will be of $A$- (non-degenerate) or $E$-type (doubly-degenerate) symmetry.
We find a non-degenerate mode with $A_{1}$-character at 218 cm$^{-1}$ whose eigenvector shows the displacement of the Xe atom to be in the $\left[ 1 1 1\right]$ direction. This displacement compresses three of the C--Xe bonds (\textit{e.g.} C$_{1-3}$) and stretches the other three.  A doubly-degenerate mode of $E$-character was found at 251 cm$^{-1}$ which is more prominent in the full vDOS than the $A_{1}$ mode.  All three modes are mutually orthogonal (see Fig. \ref{fig:vec}), and are both IR- and Raman-active.  For all three modes the relative displacement of any C atom is much smaller (by an order of magnitude) than the displacement of the Xe atom, showing that the vibrations are indeed almost entirely localized on the Xe atom.

\begin{figure}[ht]
\mbox{\subfloat{\label{fig:Xe3VvdosBZ}\includegraphics[width=0.9\linewidth]{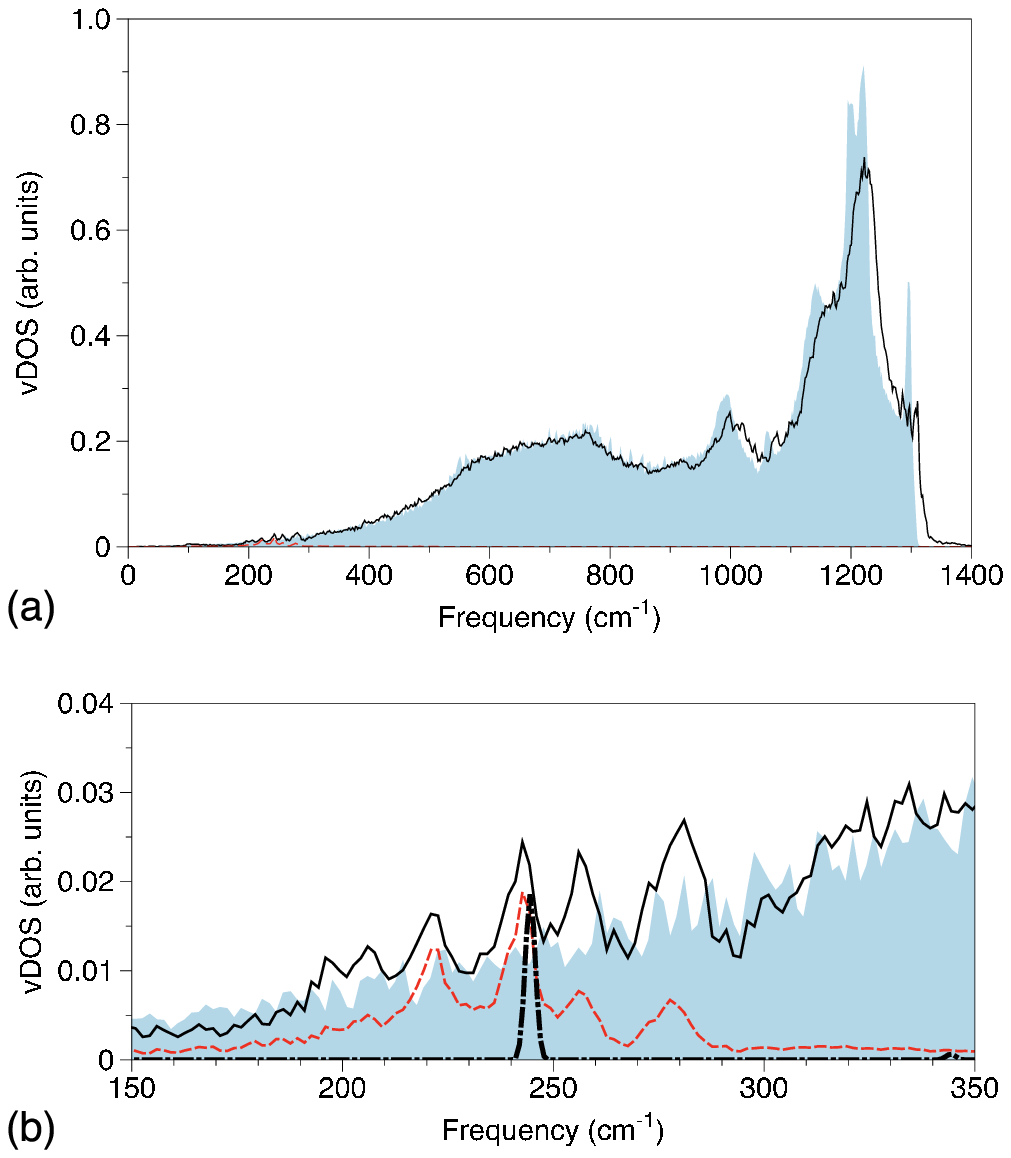}}}
\caption{(Color online)  Vibrational density of states for the relaxed Xe--3V defect.  Total vDOS (solid black line), Xe-projected vDOS (dashed red line), Gaussian-smeared Xe-projected vDOS at $\Gamma$ (dot-dashed green line), bulk diamond vDOS (shaded background).  (a) Full calculated vDOS.  (b) Region where Xe-projected vDOS is significant.}
\label{fig:Xe3VvDOS}
\end{figure}

The corresponding total vDOS of the Xe--3V defect (black) and for bulk diamond (shaded) is shown in Fig. \ref{fig:Xe3VvdosBZ} (again averaged across the Brillouin zone).  Again, the region where the Xe-projected DOS (dashed red) is significant is displayed in Fig. 7b.  At the $\Gamma$ point we find a non-degenerate mode at 245cm$^{-1}$ (dot-dashed green) which displaces the Xe atom along the $\left[ \bar{1} 1 1\right]$ direction (along the Xe--C$_{1}$ bond).  The relative displacement of any C atom is again smaller (by an order of magnitude) than the displacement of the Xe atom.  As this mode is non-degenerate, it corresponds to an $A_{1}$-type mode which is both IR- and Raman-active.

From 400 cm$^{-1}$, both vDOS spectra are largely unchanged from the bulk, apart from some softening of peaks and a slight shift from about 900 cm$^{-1}$ onwards.  In particular, the sharp peak near 1300 cm$^{-1}$ is significantly lower.

The predicted Raman signals should be visible and differentiable at sufficient doping density, especially if appropriate line-fitting models are used \cite{Asthana92}.  Some indications of them are present in the PL phonon sidebands seen in Martinovich \textit{et al.} \cite{Martinovich03}, but a clearer signal could be obtained by looking at the sidebands nearest the pump laser frequency.  This may reduce any frequency-shift effects of the electronic level excitation/de-excitation.  Also, if the center in question happens to be the optically active one, pumping its optical emission could set up a dipole which may enhance the Raman signal, making detection easier.  Unfortunately, none of the currently published experimental work on Xe centers provide spectra sufficiently close to the pump laser frequency to compare with \cite{Martinovich03,Martinovich04,Zaitsev06,Bergman07,Bergman09}.
\section{Conclusions}
\label{sec:conclusions}
We have presented the results of Density Functional Theory simulations of Xe in diamond and draw several conclusions.  First, the split-vacancy neutral defect is the most stable of those considered, at all temperatures.  Second, the cost of removing two extra C atoms from the lattice outweighs the vibrational benefit of the extra space.  Finally, both defects have IR- or Raman-active vibrational modes at $\Gamma$ which may aid in further identification of the structure of the defects studied by Bergman \textit{et al.} \cite{Bergman07, Bergman09}.
\section*{Acknowledgements}
The authors would like to thank P.G. Spizziri for several fruitful discussions, and acknowledge the support of the Australian Research Council (ARC), the National Computational Infrastructure National Facility (NCI-NF) and the Victorian Partnership for Advanced Computing (VPAC) in this work.
\bibliographystyle{unsrt}

\begin{thebibliography}{10}

\bibitem{Davies94}
G.~Davies.
\newblock {\em Properties and growth of diamond}.
\newblock EMIS Data Review Series. INSPEC, The Institution of Electrical
  Engineers, London, 1994.

\bibitem{Zaitsev00}
A.~M.~Zaitsev.
\newblock {\em Phys. Rev. B}, 61(19):12909--12922, 2000.

\bibitem{Zaitsev01}
A.~M.~Zaitsev.
\newblock {\em Optical properties of diamond: a data handbook}.
\newblock Springer-Verlag, 2001.

\bibitem{Ekimov04}
E.~A.~Ekimov, V.~A.~Sidorov, E.~D.~Bauer, N.~N.~Mel'nik, N.~J.~Curro, J.~D.~Thompson,
  and S.~M.~Stishov.
\newblock {\em Nature}, 428:542--545, 2004.

\bibitem{Takano04}
Y.~Takano, M.~Nagao, I.~Sakaguchi, M.~Tachiki, T.~Hatano, K.~Kobayashi,
  H.~Umezawa, and H.~Kawarada.
\newblock {\em App. Phys. Lett.}, 85(14):2851--2853, 2004.

\bibitem{Zaitsev06}
A.~M.~Zaitsev, A.~A.~Bergman, A.~A.~Gorokhovsky, and M.~Huang.
\newblock {\em Phys. Status Solidi A}, 203(3):638--642, 2006.

\bibitem{Davies81}
G.~Davies.
\newblock {\em Rep. Prog. Phys.}, 44:787--830, 1981.

\bibitem{Isoya90}
J.~Isoya, H.~Kanda, and Y.~Uchida.
\newblock {\em Phys. Rev. B}, 42(16):9843--9852, 1990.

\bibitem{Chrenko73}
R.~M.~Chrenko.
\newblock {\em Phys. Rev. B}, 7(10):4560--4567, 1973.

\bibitem{Kurtsiefer00}
C.~Kurtsiefer, S.~Mayer, P.~Zarda, and H.~Weinfurter.
\newblock {\em Phys. Rev. Lett.}, 85(2):290--293, 2000.

\bibitem{Gaebel04}
T.~Gaebel, I.~Popa, A.~Gruber, M.~Domhan, F.~Jelezko, and J.~Wrachtrup.
\newblock {\em New Journal of Physics}, 6:98, 2004.

\bibitem{Rabeau05}
J.~R.~Rabeau, Y.~L.~Chin, S.~Prawer, F.~Jelezko, T.~Gaebel, and J.~Wrachtrup.
\newblock {\em App. Phys. Lett.}, 86:131926, 2005.

\bibitem{Wang06}
C.~Wang, C.~Kurtsiefer, H.~Weinfurter, and B.~Burchard.
\newblock {\em J. Phys. B}, 39:37--41, 2006.

\bibitem{Simpson09}
D.~A.~Simpson, E.~Ampem-Lassen, B.~C.~Gibson, S.~Trpkovski, F.~M.~Hossain, S.~T.~Huntington, A.~D.~Greentree, L.~C.~L.~Hollenberg, and S.~Prawer.
\newblock {\em App. Phys. Lett.}, 94:203107, 2009.

\bibitem{Aharonovich09}
I.~Aharonovich, S.~Castelletto, D.~A.~Simpson, A.~Stacey, J.~McCallum, A.~D.~Greentree, and S.~Prawer.
\newblock {\em Nano Letters}, 9(9):3191--3195, 2009.

\bibitem{Jelezko04}
F.~Jelezko, T.~Gaebel, I.~Popa, M.~Domhan, A.~Gruber, and J.~Wrachtrup.
\newblock {\em Phys. Rev. Lett.}, 93(13):130501, 2004.

\bibitem{Chernobrod05}
B.~M.~Chernobrod and G.~P.~Berman.
\newblock {\em J. App. Phys.}, 97:014903, 2005.

\bibitem{Balasubramanian08}
G.~Balasubramanian, I.~Y.~Chan, R.~Kolesov, M.~Al-Hmoud, J.~Tisler, C.~Shin,
  C.~Kim, A.~Wojcik, P.~R.~Hemmer, A.~Krueger, T.~Hanke, A.~Leitenstorfer,
  R.~Bratschitsch, F.~Jelezko, and J.~Wrachtrup.
\newblock {\em Nature}, 455:648--651, 2008.

\bibitem{Maze08}
J.~R.~Maze, P.~L.~Stanwix, J.~S.~Hodges, S.~Hong, J.~M.~Taylor, P.~Cappellaro,
  L.~Jiang, M.~V.~Gurudev~Dutt, E.~Togan, A.~S.~Zibrov, A.~Yacoby, R.~L.~
  Walsworth, and M.~D.~Lukin.
\newblock {\em Nature}, 455:644--647, 2008.

\bibitem{Balasubramanian09}
G.~Balasubramanian, P.~Neumann, D.~Twitchen, M.~Markham, R.~Kolesov,
  N.~Mizuochi, J.~Isoya, J.~Achard, J.~Beck, J.~Tissler, V.~Jacques, P~Hemmer,
  F.~Jelezko, and J.~Wrachtrup.
\newblock {\em Nature Materials}, 8:383--387, 2009.

\bibitem{Cole09}
J.~H.~Cole and L.~C.~L.~Hollenberg.
\newblock {\em Nanotechnology}, 20:495401, 2009.

\bibitem{Hall09}
L.~T.~Hall, J.~H.~Cole, C.~D.~Hill, and L.~C.~L.~Hollenberg.
\newblock {\em Phys. Rev. Lett.}, 103:220802, 2009/

\bibitem{Beveratos02}
A.~Beveratos, R.~Brouri, T.~Gacoin, A.~Villing, J.-P. Poizat, and P.~Grangier.
\newblock {\em Phys. Rev. Lett.}, 89(18):187901, 2002.

\bibitem{Bennett84}
C.~H.~Bennett and G.~Brassard.
\newblock {\em Proc. IEEE International Conference on Computers,
  Systems and Signal Processing}, pages 175--179, 1984.

\bibitem{Martinovich03}
V.~A.~Martinovich, A.~V.~Turukhin, A.~M.~Zaitsev, and A.~A.~Gorokhovsky.
\newblock {\em J. Lumin.}, 102-103:785--790, 2003.

\bibitem{Martinovich04}
V.~A.~Martinovich and A.~A.~Gorokhovsky.
\newblock {\em J. Lumin.}, 107:261--265, 2004.

\bibitem{Bergman07}
A.~A.~Bergman, A.~M.~Zaitsev, and A.~A.~Gorokhovsky.
\newblock {\em J. Lumin.}, 125:92--96, 2007.

\bibitem{Bergman09}
A.~A.~Bergman, A.~M.~Zaitsev, M.~Huang, and A.~A.~Gorokhovsky.
\newblock {\em J. Lumin.}, 129:1524--1526, 2009.

\bibitem{Goss09}
J.~P.~Goss, R.~J.~Eyre, P.~R.~Briddon, and A.~Mainwood.
\newblock {\em Phys. Rev. B}, 80:085204, 2009.

\bibitem{Hohenberg64}
P.~Hohenberg and W.~Kohn.
\newblock {\em Phys. Rev.}, 136(3B):B864--B871, 1964.

\bibitem{Kohn65a}
W.~Kohn and L.~J.~Sham.
\newblock {\em Phys. Rev.}, 140(4A):A1133--A1138, 1965.

\bibitem{Goss96}
J.~P.~Goss, R.~Jones, S.~J.~Breuer, P.~R.~Briddon, and S.~\"Oberg.
\newblock {\em Phys. Rev. Lett.}, 77(14):3041--3044, 1996.

\bibitem{Schultz06}
P.~A.~Schultz.
\newblock {\em Phys. Rev. Lett.}, 96:246401, 2006.

\bibitem{Hossain08}
F.~M.~Hossain, M.~W.~Doherty, H.~F.~Wilson, and L.~C.~L.~Hollenberg.
\newblock {\em Phys. Rev. Lett.}, 101:226403, 2008.

\bibitem{Barnard03a}
A.~S.~Barnard, S.~P.~Russo, and I.~K.~Snook.
\newblock {\em Philos. Mag. Lett.}, 83(1):39--45, 2003.

\bibitem{Barnard03c}
A.~S.~Barnard, S.~P.~Russo, and I.~K.~Snook.
\newblock {\em Philos. Mag.}, 83(9):1163--1174, 2003.

\bibitem{Barnard03d}
A.~S.~Barnard, S.~P.~Russo, and I.~K.~Snook.
\newblock {\em J. Chem. Phys.}, 118(23):10725--10728, 2003.

\bibitem{Barnard03g}
A.~S.~Barnard, S.~P.~Russo, and I.~K.~Snook.
\newblock {\em Phys. Rev. B}, 68(23):235407, 2003.

\bibitem{Jelezko06}
F.~Jelezko and J.~Wrachtrup.
\newblock {\em Phys. Status Solidi A}, 203(13):3207--3225, 2006.

\bibitem{Goss07}
J.~P.~Goss, M.~J.~Shaw, and P.~R.~Briddon.
\newblock {\em Theory of Defects in Semiconductors}, volume 104 of {\em Topics
  in Applied Physics}.
\newblock Springer, Berlin/Heidelberg, 2007.

\bibitem{Kresse96}
G.~Kresse and J.~Furthm\"uller.
\newblock {\em Phys. Rev. B}, 54(16):11169--11186, 1996.

\bibitem{Perdew96}
J.~P.~Perdew, K.~Burke, and M.~Ernzerhof.
\newblock {\em Phys. Rev. Lett.}, 77(18):3865--3868, 1996.

\bibitem{Kresse99}
G.~Kresse and D.~Joubert.
\newblock {\em Phys. Rev. B}, 59(3):1758--1775, 1999.

\bibitem{Barnard02}
A.~S.~Barnard and S.~P.~Russo and I.~K.~Snook.
\newblock {\em Philos. Mag. B}, 82(17):1767--1776, 2002.

\bibitem{Barnard03h}
A.S. Barnard.
\newblock {\em Ab Initio Modelling of Nanocrystalline Diamond in 0 and 1
  Dimensions}.
\newblock PhD thesis, RMIT, 2003.

\bibitem{Alfe09}
D.~Alf\`e.
\newblock {\em Comput. Phys. Commun.}, 180:2622--2633, 2009.

\bibitem{Asthana92}
B.~P.~Asthana and W.~Keifer, in
\newblock {\em Vibrational spectra and structure}, chapter ``Vibrational line
  profile and frequency shift studies by Raman spectroscopy'', edited by J.~R.~Durig, pages 67--155.
\newblock Elsevier Science Publishers, 1992.

\end{thebibliography}

\end{document}